\documentclass[a4paper,fleqn,usenatbib]{mnras}

\usepackage{mathptmx}

\usepackage[T1]{fontenc}
\usepackage{ae,aecompl}

\usepackage{graphicx}	
\usepackage{amsmath}	
\usepackage{amssymb}	
\usepackage{booktabs}

\newcommand{\Msun}{\,\mathrm{M_{\sun}}}

\title[]{Connecting optical and X-ray tracers of galaxy cluster relaxation}

\author[I.D. Roberts et al.]{
Ian D. Roberts$^{1}$\thanks{E-mail: roberid@mcmaster.ca},
Laura C. Parker$^{1}$, Julie Hlavacek-Larrondo$^{2}$
\\
$^{1}$Department of Physics and Astronomy, McMaster University, Hamilton ON
L8S 4M1, Canada \\
$^{2}$D\'{e}partement de Physique, Universit\'{e} de Montr\'{e}al, Montr\'{e}al QC H3C 3J7, Canada \\
}

\date{Accepted XXX. Received YYY; in original form ZZZ}

\pubyear{2017}

\begin{document}
\label{firstpage}
\pagerange{\pageref{firstpage}--\pageref{lastpage}}
\maketitle

\begin{abstract}
  Substantial effort has been devoted in determining the ideal proxy for quantifying the morphology of the hot intracluster medium in clusters of galaxies.  These proxies, based on X-ray emission, typically require expensive, high-quality X-ray observations making them difficult to apply to large surveys of groups and clusters.  Here, we compare optical relaxation proxies with X-ray asymmetries and centroid shifts for a sample of SDSS clusters with high-quality, archival X-ray data from \textit{Chandra} and \textit{XMM-Newton}.  The three optical relaxation measures considered are: the shape of the member-galaxy projected velocity distribution -- measured by the Anderson-Darling (AD) statistic, the stellar mass gap between the most-massive and second-most-massive cluster galaxy, and the offset between the most-massive galaxy (MMG) position and the luminosity-weighted cluster centre.  The AD statistic and stellar mass gap correlate significantly with X-ray relaxation proxies, with the AD statistic being the stronger correlator.  Conversely, we find no evidence for a correlation between X-ray asymmetry or centroid shift and the MMG offset.  High-mass clusters ($M_\mathrm{halo} > 10^{14.5}\Msun$) in this sample have X-ray asymmetries, centroid shifts, and Anderson-Darling statistics which are systematically larger than for low-mass systems.  Finally, considering the dichotomy of Gaussian and non-Gaussian clusters (measured by the AD test), we show that the probability of being a non-Gaussian cluster correlates significantly with X-ray asymmetry but only shows a marginal correlation with centroid shift.  These results confirm the shape of the radial velocity distribution as a useful proxy for cluster relaxation, which can then be applied to large redshift surveys lacking extensive X-ray coverage.
\end{abstract}




\section{Introduction}
\label{sec:introduction}

\par
The majority of galaxies in the local Universe do not evolve in isolation but instead inhabit dense environments such as groups and clusters \citep[e.g.][]{geller1983, eke2005}.  In addition to internal processes (e.g.\ AGN feedback, \citealt{dubois2013, gurkan2015, mullaney2015, bongiorno2016}; bar-driven evolution, \citealt{knapen1995, kormendy2004, sheth2005}; morphological quenching, \citealt{martig2009}; virial gas heating, \citealt{birnboim2003, cattaneo2006, gabor2015}; etc.), interactions with local environments play a significant role in shaping the observed properties of galaxies.  For example, mechanisms acting in dense environments such as ram-pressure stripping \citep[e.g.][]{gunn1972} and starvation \citep[e.g.][]{larson1980, peng2015} can remove the cold and hot gas components from galaxies, respectively.  Galaxy interactions, such as mergers and impulsive high-speed encounters, can drive gas to the central regions and induce star-burst events which may exhaust a galaxies gas reserves \citep[e.g.][]{mihos1994a, mihos1994b, ellison2008, davies2015}.  These interactions can also influence galaxy morphology through the growth of a strong bulge component, and the end products of major mergers tend to be bulge dominated galaxies with classical de Vaucouleurs profiles \citep[e.g.][]{barnes1989}.  Finally, tidal interactions can also influence gas content through direct stripping or by transporting gas outwards allowing it to be more easily stripped by other mechanisms \citep[e.g.][]{mayer2006, chung2007}. It's generally accepted that these mechanisms can act on galaxies in dense environments, though the relative balance between different mechanisms in different environments remains an outstanding question.
\par
Understanding the influence of environment is contingent on being able to identify and quantify galaxy environments.  Common environmental measures include the projected number density of galaxies out to the $Nth$ nearest neighbour, the halo mass of a host group or cluster, or the projected separation from the centre of a group or cluster.  Star formation and morphology of galaxies correlate well with these environment proxies, with galaxies in high densities regions (or alternatively, high halo mass or small group/cluster-centric radius) being preferentially red, passive, and early type \citep{dressler1980, goto2003, poggianti2008, kimm2009, li2009, wetzel2012, wilman2012, fasano2015, haines2015}.
An alternative way to parametrize the environment of a host group or cluster, is to classify the degree to which a system is dynamically relaxed.  A relaxed, dynamically old group or cluster should be characterized by a central galaxy which is the brightest (most massive) member by a significant margin \citep[e.g][]{khosroshahi2007, dariush2010, smith2010} and is located near the minimum of the potential well (e.g.\ \citealt{george2012, zitrin2012}, however also see \citealt{skibba2011}), satellite galaxies which are distributed in velocity space according to a Gaussian profile \citep[e.g.][]{yahil1977, bird1993, hou2009, martinez2012}, and diffuse X-ray emission which is symmetric about the group/cluster centre \citep[e.g.][]{rasia2013, weissmann2013, parekh2015}.  The dynamical state of clusters is related to the age of the halo and the time since infall for member galaxies, which simulations have shown is an important quantity in determining the degree to which galaxy properties are affected by environment \citep[e.g.][]{wetzel2013, oman2016, joshi2017}.  Unrelaxed groups and clusters are systems which formed more recently or which have recently experienced a significant merger event, and in either case it would be expected that the time-since-infall onto the current halo for member galaxies will be relatively short.  Therefore galaxies in unrelaxed groups may have properties which have been less influenced by environment compared to galaxies in more relaxed systems.
\par
Recent studies have attempted to determine the degree to which galaxy properties depend on the ``relaxedness'' of a given group or cluster.  It has been shown that galaxies in relaxed groups tend to be redder than counterparts in unrelaxed systems, using relaxation definitions based on the presence of a well-defined central galaxy \citep[e.g.][]{carollo2013} as well as the shape of the satellite velocity distribution \citep[e.g.][]{ribeiro2010, ribeiro2013a}.  Previously, we have shown that low-mass galaxies in the inner regions of Gaussian (G) groups have reduced star-forming fractions relative to non-Gaussian (NG) groups \citep{roberts2017}.  We have also shown that star-forming and disc fractions for low mass galaxies are enhanced in X-ray underluminous (XRW) groups, and show that galaxies XRW groups have velocity distributions consistent with being unrelaxed systems (at least relative to X-ray strong groups, \citealt{roberts2016}).
\par
Building from our recent work, here we aim to further investigate the connection between X-ray and optical measures of group relaxedness.  The shape of the diffuse X-ray component of a group or cluster is among the most direct probes of the degree to which a group/cluster is relaxed or recently disturbed.  The downside, however, is that measuring this morphology requires deep, high-quality X-ray observations which are not available for large surveys containing thousands of groups and clusters.  To address this challenge, we use a sample of galaxy clusters with existing X-ray observations to investigate the relationship between the X-ray relaxation and three previously used optical probes of relaxation: the shape of the satellite velocity distribution, the stellar mass gap between the most-massive and second-most-massive group galaxy, and the offset between the position of the most-massive galaxy and the luminosity-weighted centre of the group.  We determine the effectiveness of these optical relaxation measures (which are applicable to large redshift surveys) by comparing them to measured X-ray morphology, a more direct probe of relaxation.
\par
The outline of this paper is as follows.  In Section 2 we describe the optical group catalogue as well as the archival X-ray data used in this work.  In Section 3 we outline the cluster relaxation estimators, both optical and X-ray, that we consider.  In Section 4 we present the main results, comparing optical and X-ray cluster relaxation measures.  In Section 5 we discuss these results and provide a summary in Section 6.
\par
This paper assumes a flat $\mathrm{\Lambda}$ cold dark matter cosmology with $\Omega_\mathrm{M} = 0.3$, $\Omega_\mathrm{\Lambda} = 0.7$, and $H_0 = 70\,\mathrm{km}\,\mathrm{s^{-1}}\,\mathrm{Mpc^{-1}}$.  The $h$-dependence of important calculated properties are: $M_\mathrm{halo} \sim h^{-1}$, $M_\mathrm{\star} \sim h^{-2}$, $R_{500} \sim h^{-1}$.


\section{Data}
\label{sec:data}

\begin{figure}
	\centering
	\includegraphics[width=0.85\columnwidth]{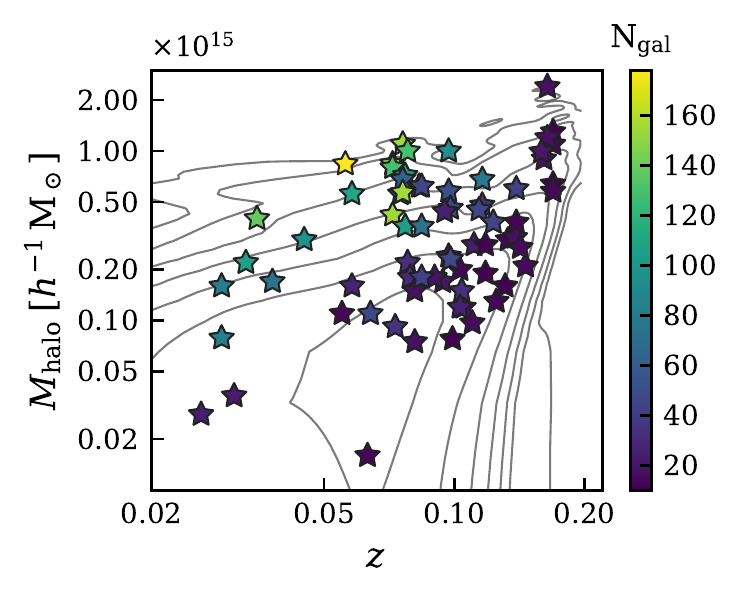}
	\caption{Cluster halo mass versus redshift for Yang clusters.  Stars correspond to the X-ray matched clusters used in this work, coloured by the number of galaxies identified in each system.  Gray contours show the distribution for the parent sample of N > 10 Yang clusters.}
	\label{fig:z_mh}
\end{figure}

\subsection{Optically identified galaxy clusters}

We use galaxy clusters identified from the seventh release of the Sloan Digital Sky Survey (SDSS DR7; \citealt{abazajian2009}) by \citet{yang2005, yang2007} who construct a group sample using a ``halo-based" group finder which aims to improve upon the classic friends-of-friends (FoF) algorithm \citep[e.g.][]{huchra1982, press1982}.  For a full description of the algorithm see \citet{yang2005, yang2007}, however in short, the groups are initially populated by connecting galaxies through a standard FoF approach (with very small linking lengths) and group memberships are iteritively updated under the assumption that the distribution of galaxies in phase space follows that of a spherical NFW profile \citep{navarro1997}.  Each iteration yields an updated estimate of the group mass, size, and velocity dispersion and iterations continue until memberships stabilize.  Final group halo masses ($M_\mathrm{halo}$) obtained via abundance matching are given in the Yang catalogue (in particular, we use the sample III); we use galaxy stellar masses ($M_\star$) given in the New York University Value-Added Galaxy Catalogue (NYU-VAGC; \citealt{blanton2005a}) determined using fits to the galaxy spectra and broad-band photometric measurements following the procedure of \citet{blanton2007}.  We note that the Yang catalogue contains a mixture of what would generally be considered groups ($M_\mathrm{halo} < 10^{14}\Msun$) as well as galaxy clusters ($M_\mathrm{halo} \ge 10^{14}\Msun$), for the sake of brevity we will refer to all systems as clusters regardless of halo mass as the majority of the systems we consider have $M_\mathrm{halo} \ge 10^{14}\Msun$.
\par
Cluster-centric radii are computed for galaxies using the redshift and the angular separation between the galaxy position and the luminosity-weighted centre of the cluster.  We normalize all cluster-centric radii by $R_{500}$ (the radius at which the average interior density is 500 times the critical density of the Universe) of each cluster which we compute as
\begin{equation}
	R_{500} = R_{200m} / 2.7,
\end{equation}
\noindent
where,
\begin{equation}
	R_{200m} = 1.61\,\mathrm{Mpc} \left(\frac{M_\mathrm{halo}}{10^{14}\,\Msun}\right)^{1/3} (1 + z_\mathrm{group})^{-1}
\end{equation}
\noindent
is the radius at which the average interior density is equal to 200 times the critical mass density of the Universe \citep{yang2007, tinker2008}, and we have
assumed an NFW density profile \citep{navarro1997} with a concentration given by the concentration-mass relation of \citet{maccio2007} \citep{wang2014}.  

Our sample of galaxy clusters is a subset of the Yang catalogue including only clusters with ten or more member galaxies (2559 clusters).  The cut-off in membership is chosen in order to be able to classify the shape of the velocity profile for each cluster with relative accuracy \citep{hou2009}.  Fig.~\ref{fig:z_mh} shows the $M_\mathrm{halo}$ - redshift distribution for the parent sample (gray contours) with the 58 X-ray matched clusters (see Section~\ref{sec:X-ray_match}) overplotted as stars colour-coded by the number of galaxies identified in each cluster.  As expected, at fixed redshift the observed cluster richness increases with halo mass and at fixed halo mass the observed cluster richness decreases with redshift.  The latter is a selection effect due to increasing incompleteness at higher redshift.  To check whether this incompleteness may be biasing our results we repeat our analysis on ``low-z'' ($z < 0.10$) and ``high-z'' ($z \ge 0.10$) subsamples (results not shown) and find no difference between the conclusions drawn from either redshift subsample.  Therefore moving forward, we consider the entire redshift range.

\subsection{X-ray matched clusters}
\label{sec:X-ray_match}

\begin{figure}
	\centering
	\includegraphics[width=\columnwidth]{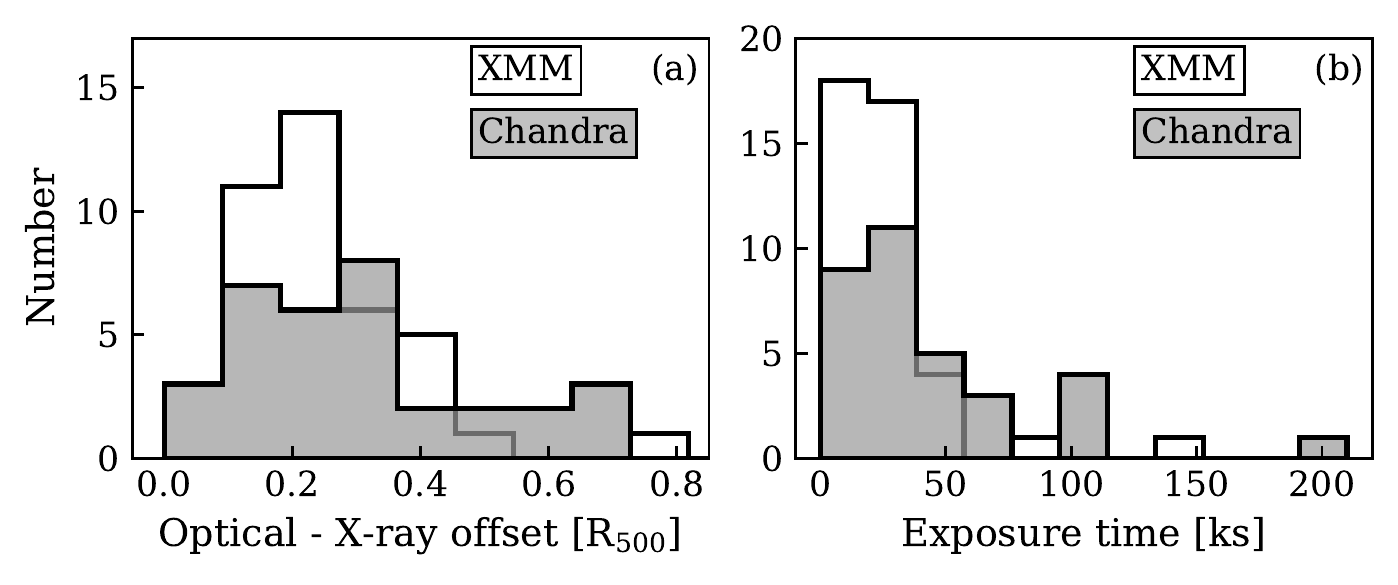}
	\caption{Distribution of the offset between the X-ray peak and cluster luminosity-weighted centre (left) and X-ray exposure time (right) for the clusters in our sample, for \textit{Chandra} (gray) and \textit{XMM-Newton} (white).}
	\label{fig:offset_texp}
\end{figure}

In order to make connections between optical measures of cluster relaxedness and the shape of the cluster extended X-ray profile we searched the \textit{Chandra} and \textit{XMM-Newton} science archives at the positions of the luminosity-weighted centres of each of the 2559 $N > 10$ Yang clusters.  Using a search radius of $5\arcmin$ we matched observations of extended X-ray emission to the corresponding optically identified cluster, only including observations with clean exposure times $\ge 10\,\mathrm{ks}$.  We also exclude systems where multiple Yang (N > 10) clusters are matched to the same X-ray observation to avoid the potential overlap of X-ray emission from physically distinct systems in projection (this was only the case for $<5$ per cent of matches).  This matching results in 58 Yang clusters with X-ray coverage.  Fig.~\ref{fig:offset_texp}a shows the projected separation between the luminosity-weighted centre and the X-ray centre of each Yang cluster for \textit{Chandra} (gray) and \textit{XMM-Newton} (white), whereas Fig.~\ref{fig:offset_texp}b shows the respective filtered exposure times for the observations.  X-ray centres are calculated as the position of the brightest pixel in the X-ray image after smoothing using a Gaussian kernel with a bandwidth of $40\,\mathrm{kpc}$ (as in \citealt{nurgaliev2013}).  As shown in Fig.~\ref{fig:offset_texp}a, the offset between the optical and X-ray centres is far smaller than the virial radius for all systems.
\par
\textit{Chandra} observations were reprocessed, cleaned, and calibrated using the latest version of \textsc{ciao} (\textsc{ciao} version 4.9, \textsc{caldb} version 4.7.5).  Charge transfer inefficiency and time-dependent gain corrections were applied and observations were filtered for background flares using the \textsc{lc\_clean} script with a $3\sigma$ threshold.  Exposure corrected images are then created using exposure maps generated at an energy of $1.5\,\mathrm{keV}$, the average peak emission of our sample.  Images were created in the $0.5-5\,\mathrm{keV}$ energy band to maximize the ratio between cluster and background flux \citep{nurgaliev2013}.  Point sources are identified using the \textsc{wavdetect} script and are filled with local poisson noise using \textsc{dmfilth}, blank sky background images are generated for each observation using the \textsc{blanksky} and \textsc{blanksky\_image} scripts.  All observations are then checked by eye to ensure that no obvious point sources were missed by the algorithm.  For systems with multiple observations, combined images and exposure maps were generated with the \textsc{merge\_obs} script and blank sky background images were combined using \textsc{reproject\_image}. 
\par
Data reduction for \textit{XMM-Newton} observations was done using the Extended Source Analysis Software (\textsc{esas}) within the \textit{XMM-Newton} Science Analysis System (\textsc{sas}, version 16.0.0).  Calibrated event files were generated using the \textsc{emchain} script, and filtered event lists were generated using \textsc{mos-filter}.  Exposure corrected images were created in the $0.5-5\,\mathrm{keV}$ band, and point sources were identified with the \textsc{cheese} script and subsequently filled with local poisson noise using the \textsc{ciao} script \textsc{dmfilth}.  Again, images are checked by eye to ensure no obvious point sources are missed.  For \textit{XMM-Newton} observations we only use the MOS exposures to avoid the complications of the many chip gaps on the PN detector.  MOS exposures are combined using the \textsc{comb} script to give merged images, exposure maps, and background images.  For systems with multiple observations, images, exposure maps, and background images are merged with the \textsc{ciao} script \textsc{reproject\_image}.
\par
The pixel scale of the resulting images is $0.5\arcsec$ for \textit{Chandra} and $2.5\arcsec$ for \textit{XMM-Newton}.  We calculate X-ray asymmetries and centroid shifts at these native resolutions to avoid losing information from the higher resolution \textit{Chandra} images, however we note that binning the \textit{Chandra} images to the \textit{XMM-Newton} resolution does not alter the results.  Furthermore, when we compare asymmetries and centroid shifts computed for systems which are observed by both \textit{Chandra} and \textit{XMM-Newton}, we see no bias introduced by the resolution difference.

\section{Cluster relaxation measures}
\label{sec:relax_param}

\subsection{Optical}
In this study we implement three previously used optical measures to parameterize the relaxation of clusters: the Anderson-Darling statistic, the stellar mass ratio between the second most-massive and the most-massive cluster galaxy ($M_2 / M_1$), and the offset between the position of the MMG and the luminosity-weighted centre of the cluster (MMG offset).

\subsubsection{Anderson-Darling statistic}

The Anderson-Darling (AD) test is a statistical normality test which measures the ``distance" between the cumulative distribution functions (CDFs) corresponding to the data as well as the ideal case of a normal distribution \citep{anderson1952}.  The distance between the CDFs is parameterized by the AD statistic ($A^2$), in the sense that large values of this statistic correspond to larger deviations from normality.  The AD statistic, $A^2$ is given by
\begin{equation}
  A^2 = -n - \frac{1}{n} \sum_{i=1}^n [2i - 1][\ln \Phi(x_i) + \ln(1 - \Phi(x_{n+1-i}))],
\end{equation}
\noindent
where $x_i$ are the length-$n$ ordered data and $\Phi(x_i)$ is the CDF of the hypothetical underlying distribution (Gaussian in this application).
\par
In the context of cluster evolution, it is expected that galaxies in evolved, dynamically old clusters should display projected velocity profiles which are well fit by a normal distribution; conversely more unrelaxed clusters will show larger deviations from normality \citep{yahil1977, bird1993, ribeiro2013a}. The AD test can therefore be applied to the velocity distributions of member galaxies to discriminate between relaxed and unrelaxed clusters \citep[e.g.][]{hou2009}.  In this work we use the AD statistic as a proxy for cluster relaxedness, where increasing values of $A^2$ are indicative of progressively more unrelaxed clusters.  It is also common in the literature to use the p-value associated with the AD statistic to define a dichotomy between Gaussian and non-Gaussian clusters \citep[e.g][]{hou2009, martinez2012, roberts2017}, which we consider in Section~\ref{sec:G_NG}.

\subsubsection{Stellar mass gap}

The second optical parameter we use to classify the relaxation of galaxy clusters is the stellar mass ratio between the second most-massive and most-massive galaxies in a given cluster.  Since the MMG should sit near the centre of the cluster potential, it will progressively grow in stellar mass by dominating gas accretion within the cluster, and more importantly, by cannibalizing galaxies through minor mergers \citep[e.g.][]{delucia2007, ruszkowski2009,lin2013, mcdonald2016}.  This MMG mass growth will therefore drive down $M_2/M_1$ in dynamically old clusters, whereas more unrelaxed systems will have had less time to establish a dominant MMG.
\par
The reliability of $M_2/M_1$ is contingent on correctly identifying both the MMG and $M_2$.  A particular concern when using SDSS data is the potential for galaxies missing spectra due to fibre collisions; this has an increasing impact in the dense inner regions of groups and clusters where one would expect to find the MMG and $M_2$.  In an attempt to mitigate the effects of fibre collisions we use sample III from the Yang group catalogue which corrects for fibre collisions by assigning fibre collision galaxies the redshift of the galaxy they ``collide'' with.  While this procedure accounts for fibre collisions it also introduces potential impurities to the group catalogue (some fibre collision galaxies will have redshifts which are catastrophically different from the one they are assigned), we delay a more detailed discussion of these effects until Section~\ref{sec:mmg_discussion} though we urge the reader to keep these caveats in mind when interpretting results in Section~\ref{sec:xray_opt_corr}.

\subsubsection{MMG offset}

The final optical relaxation parameter we consider is the projected offset between the MMG and the luminosity-weighted cluster centre, $\delta R_{MMG}$.  There is currently no consensus regarding the best observational definition of group centre, with the position of the MMG, the position of the X-ray peak, and the luminosity or mass-weighted centre all being popular choices \citep[e.g.][]{george2012}.  For relaxed clusters it is expected that all of the aforementioned centre definitions will be relatively consistent with one another, however more unrelaxed clusters may show significant offsets between different cluster centre choices.  In particular, many unrelaxed clusters host MMGs with large offsets from other cluster centre definitions \citep[e.g.][]{katayama2003, sanderson2009, carollo2013, khosroshahi2017}, therefore the offset between MMG position and luminosity-weighted centre can be a useful measure of cluster relaxation.
\par
As with the stellar mass gap, there are potential complications with regards to interpretting the MMG offset as a relaxation probe.  For example, it is based on the assumption that in a relatively relaxed systems the MMG (or brightest galaxy) will be located at rest at the centre of the dark matter potential well -- the so-called central galaxy paradigm (CGP).  However, some recent studies have called into question whether or not the CGP is valid in all systems \citep{vandenbosch2005, coziol2009, skibba2011, sehgal2013, lauer2014, hoshino2015}.  Additionally, even in relaxed systems the MMG may oscillate about the centre of a cored dark-matter potential \citep[e.g.][]{harvey2017} further complicating the interpretation of the radial offset of the central galaxy.  Yet again, we will defer a full discussion of these effects to Section~\ref{sec:mmg_discussion}.

\subsection{X-ray}

\begin{figure*}
	\centering
	\includegraphics[width=0.8\textwidth]{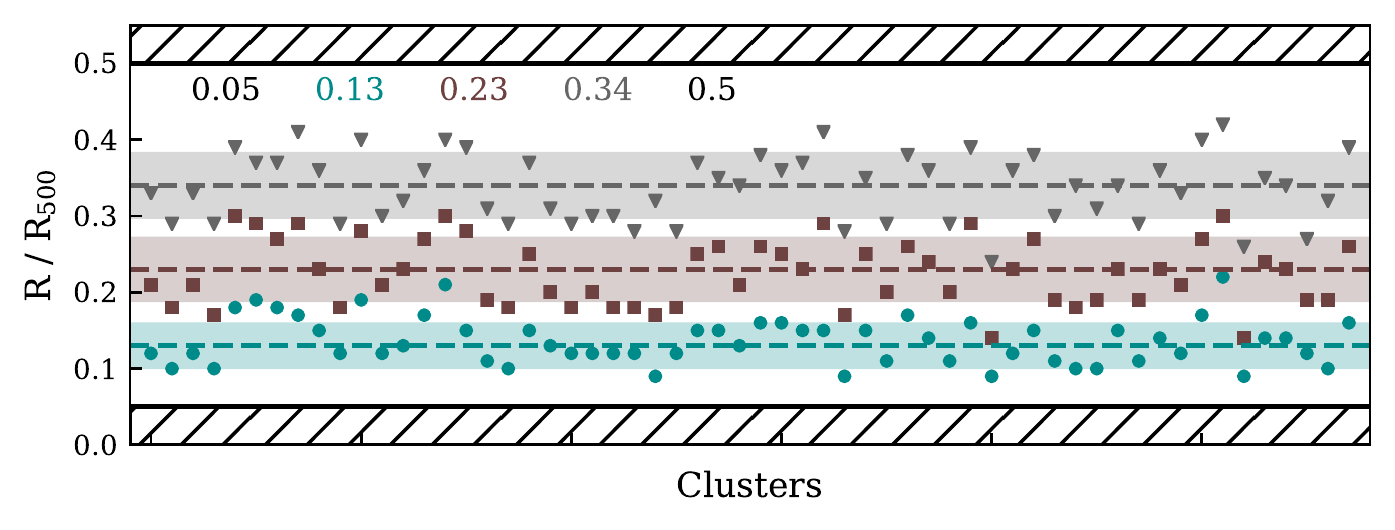}
	\caption{``Optimal annuli" positions (constant cluster counts) for each cluster in the X-ray matched sample.  For four annuli, this amounts to two fixed end points (black) and three inner boundaries with variable positions (teal circles, maroon squares, gray triangles).  Dashed lines correspond to the median value for each annulus boundary (values of the boundaries are also printed at the top of the figure), which we take to be our final annuli positions when computing asymmetries.  Hatched areas denote radial regions not included in the aymmetry calculation (see Section~\ref{sec:photon_asym}) and shaded regions show the $1\sigma$ scatter for each annuli position.}
	\label{fig:opt_ann}
\end{figure*}

To measure the degree of cluster relaxedness from X-ray observations we consider two relaxation proxies: the photon asymmetry ($A_\mathrm{phot}$) and the centroid shift ($w$).

\subsubsection{Photon Asymmetry}
\label{sec:photon_asym}

Photon asymmetry is a novel technique to measure the asymmetry of X-ray profiles which is model-independent and robust across a wide range in X-ray counts and background level \citep{nurgaliev2013}.  In this work we will give a brief discussion of the photon asymmetry computation, however for a complete description, including tests of robustness, we direct the reader to \citet{nurgaliev2013}.
\par
The photon asymmetry measures the degree to which the count profile of an X-ray observation is axisymmetric around the X-ray peak, or phrased alternatively, the degree to which the polar angles of X-ray counts are distributed uniformly over the range $0 \le \phi \le 2\pi$.  This is accomplished quantitatively using Watson's U2 test which compares the polar angle CDF for observed counts to a uniform CDF corresponding to an idealized axisymmetric profile \citep{watson1961}.  In a given radial annulus, the distance between the observed count distribution and a uniform distribution is given by
\begin{equation}
	\hat{d}_{N,C} = \frac{N}{C^2} \left(U_N^2 - \frac{1}{12} \right),
\end{equation}
\noindent
where $N$ is the total number of counts within the annulus, $C$ is the number of counts intrinsic to the cluster (ie.\ above the background) within the annulus, and $U_N^2$ is Watson's statistic.  We follow \citet{nurgaliev2013} and compute Watson's statistic using the following relation (see \citealt{watson1961})
\begin{equation}
	U_N^2(\phi_0) = \frac{1}{12N} + \sum_{i=0}^{N-1} \left(\frac{2i + 1}{2N} - F(\phi_i) \right)^2 - N \left(\frac{1}{2} - \frac{1}{N} \sum_{i=0}^{N-1} F(\phi_i) \right)^2,
\end{equation}
\noindent
where $\phi_i$ are the observed count polar angles, $\phi_0$ is the origin polar angle on the circle, and $F$ is the uniform CDF.  To obtain the final value for $U_N^2$ we minimize the statistic over all origin angles on the circle
\begin{equation}
	U_N^2 = \min_{\mathrm{origin\;on\;circle},\, \phi_0} U_N^2(\phi_0).
\end{equation}
\noindent
The final value for the photon asymmetry, $A_\mathrm{phot}$, is given by the cluster count weighted average of $\hat{d}_{N,C}$ in each radial annulus, namely
\begin{equation}
	A_\mathrm{phot} = 100 \left. \sum_{k=1}^{N_\mathrm{ann}} C_k \hat{d}_{N_k,C_k} \middle/ \sum_{k=1}^{N_\mathrm{ann}} C_k \right..
\end{equation}

We assume a uniform background which we estimate from blank-sky images for each observation and subsequently compute the number of cluster counts, $C$, by subtracting the expected number of background counts within the annulus from the total number of observed counts.  Following \citet{nurgaliev2013} we compute $\hat{d}_{N,C}$ in four radial annuli, which in this work range between $0.05\,\mathrm{R_{500}}$ and $0.5\,\mathrm{R_{500}}$.  This choice of four annuli ensures that we will obtain at least hundreds of cluster counts in each annulus for the low-count observations ($\sim$ a few thousand counts).  Optimal annuli are selected by requiring an approximately constant number of cluster counts within each annulus.  We define the annuli radii as those which minimize the variance in cluster counts across each of the annuli.  In Fig.~\ref{fig:opt_ann} we show the optimal annuli positions for each of the 58 clusters in the X-ray matched sample.  We note that while there is some variation in optimal annuli from cluster to cluster, in general the scatter is relatively small and there is no overlap between the $1\sigma$ scatter of neighbouring annuli.  The final annuli edges are taken to be the median values across all of the clusters, which corresponds to $\{0.05, 0.13, 0.23, 0.34, 0.50\}\times \mathrm{R_{500}}$.  The inner boundary of $0.05\,\mathrm{R_{500}}$ is set to aviod pixelation artifacts at small radii \citep{nurgaliev2013}, and the outer boundary of $0.5\,\mathrm{R_{500}}$ is chosen to enclose the majority of the emission while still ensuring chip coverage.  The large angular sizes of some of the high-mass, low-redshift systems ($R_{500} \sim 15 - 20 \arcmin$) prevents us from computing $A_\mathrm{phot}$ out to a full $\mathrm{R_{500}}$ since they extend beyond the edge of the detector.
\par
Statistical uncertainties on $A_\mathrm{phot}$ are estimated following \citet{nurgaliev2013} by randomly resampling half of the observed counts 500 times and recalculating $A_\mathrm{phot}$ for each iteration.  For clusters with both \textit{Chandra} and \textit{XMM-Newton} observations, we compute $A_\mathrm{phot}$ for the \textit{Chandra} and \textit{XMM} data separately and then combine them as a count-weighted average.

\subsubsection{Centroid shift}
\label{sec:centroid_shift}

A commonly used X-ray relaxation proxy is the centroid shift, $w$, which measures the shift of the X-ray surface brightness centroid in different radial apertures.  For a system in dynamical equilibrium, the centre of mass of the ICM (ie.\ the centroid) should be independent of scale, whereas an unrelaxed system with substructure can have a centre of mass which depends on radius \citep[e.g.][]{mohr1993}.  To compute centroid shifts we use the following relation \citep[e.g.][]{boehringer2010}
\begin{equation}
  w = \left[\frac{1}{N - 1}\sum_i (\Delta_i - \langle \Delta \rangle)^2 \right]^{1/2} \times \frac{1}{R_\mathrm{max}},
\end{equation}
\noindent
where $\Delta_i$ is the offset between the X-ray peak and the centroid position within the $i$th aperture, $N$ is the number of apertures, and $R_\mathrm{max}$ is the radius of the largest aperture.  Centroids are determined from the moments of the exposure-corrected X-ray images\footnote{http://photutils.readthedocs.io/en/stable/photutils/centroids.html}, and the X-ray peak is considered to be the position of the brightest pixel after smoothing using a Gaussian kernel with a bandwidth of $40\,\mathrm{kpc}$.  The smallest aperture that we consider is $R < 0.1\,\mathrm{R_{500}}$ and we progressively increase the aperture radius by $0.05\,\mathrm{R_{500}}$ out to a maximum of $0.5\,\mathrm{R_{500}}$, for a total of 9 apertures.  These aperture choices are motivated by previous studies \citep{boehringer2010, nurgaliev2013, rasia2013, weissmann2013} as well as ensuring chip coverage as was done to measure $A_\mathrm{phot}$ in Section~\ref{sec:photon_asym}.
\par
As with $A_\mathrm{phot}$, uncertainties for the centroid shift are determined from randomly resampling the X-ray images and recalculating $w$, and for clusters with observations from \textit{Chandra} and \textit{XMM-Newton} $w$ is computed as a count-weighted average.

\section{Results}
\label{sec:results}

\subsection{Relationship between X-ray and optical relaxation proxies}
\label{sec:xray_opt_corr}

To explore the consistency between group relaxation measures in the X-ray and optical, we measure the correlations between photon asymmetry and centroid shifts and the three optical relaxation parameters ($A^2$, $M_2/M_1$, $\delta R_{MMG}$).  To quantify the correlations between these parameters, we use two different methods:

\begin{enumerate}
	\item We fit a simple power-law to each relationship and derive uncertainties on the slope and normalization with bootstrap resampling.

	\item We compute Spearman's rank correlation coefficient, $r_s$, (which is preferred over the Pearson correlation due to its non-parametric nature) for each relationship to quantify the percentile at which the data are consistent with a correlation.	
\end{enumerate}

\subsubsection{Photon Asymmetry}
\label{sec:Aphot_corr}

\begin{figure*}
	\centering
	\includegraphics[width=\textwidth]{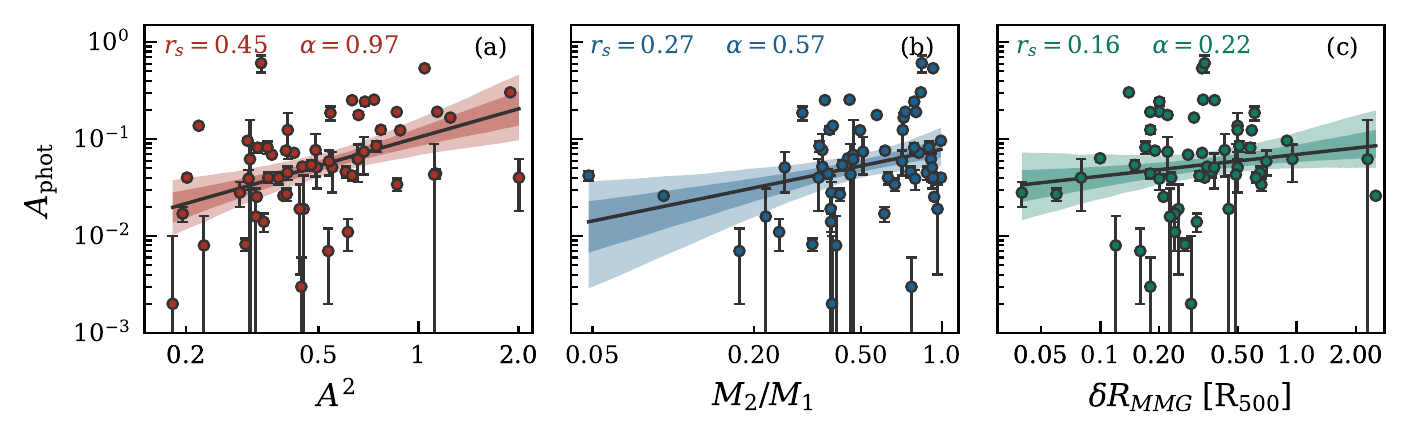}
	\caption{Photon asymmetry versus optical relaxation parameters ($A^2$, $M_2/M_1$, $\delta R_{MMG}$), error bars are $1\sigma$ resampling uncertainties.  The solid line is the best-fit power-law relationship and the shaded regions correspond to the 68 and 95 per cent bootstrap confidence intervals.  The Spearman correlation coefficient, $r_S$, and the best-fit power-law slope, $\alpha$, are indicated in the upper region of each panel.}
	\label{fig:asym_ad_m12_offset}
\end{figure*}

In Figs~\ref{fig:asym_ad_m12_offset}a-c we show the relationship between photon asymmetry and the three optical relaxation proxies.  The $A_\mathrm{phot} - A^2$ relationship shows a significant correlation as measured by both the power-law fit and by the Spearman test.  The best-fit power law has a positive slope at $3.5\sigma$ and the Spearman test gives a positive correlation at the $>99.9$ per cent level.  The $A_\mathrm{phot} - M_2 / M_1$ relationship shows a weaker (but still significant) correlation with a positive slope at $2.9\sigma$ and a Spearman correlation at the $96$ per cent level.  In contrast, the $A_\mathrm{phot} - \delta R_{MMG}$ relationship  does not display a significant correlation by either measure, with a power law slope consistent with zero and a Spearman p-value of 0.23.

\subsubsection{Centroid shift}
\label{sec:w_corr}

\begin{figure*}
	\centering
	\includegraphics[width=\textwidth]{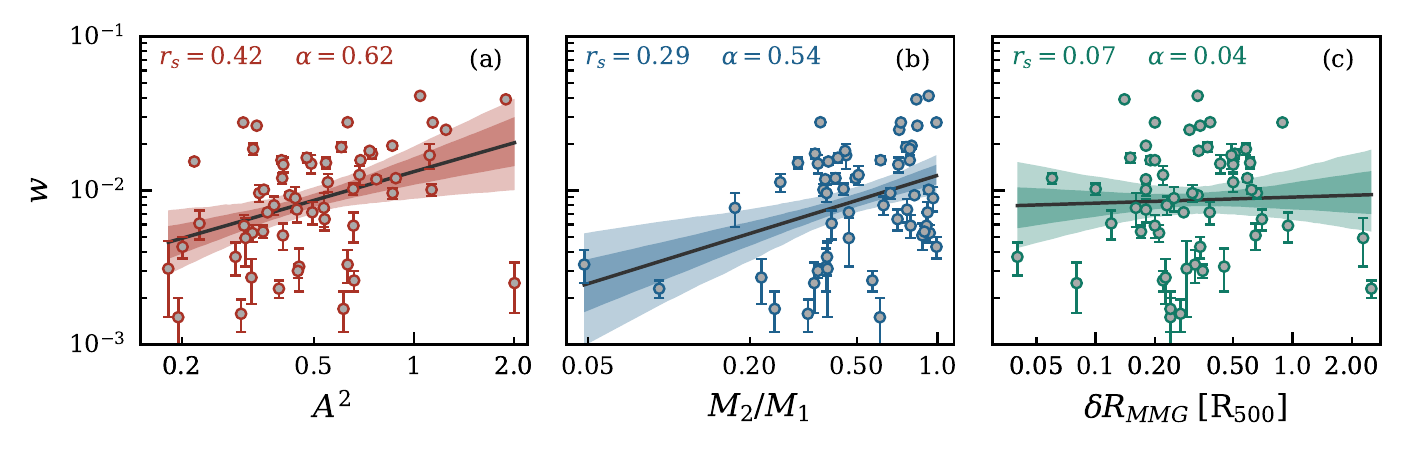}
	\caption{Same as Fig.~\ref{fig:asym_ad_m12_offset} however for the centroid shift instead of photon asymmetry.}
	\label{fig:centshift}
\end{figure*}
In Figs~\ref{fig:centshift}a-c we now show the relationship between the optical relaxation parameters and the centroid shift as the X-ray relaxation proxy.  The results in Fig.~\ref{fig:centshift} are very similar to those in Fig.~\ref{fig:asym_ad_m12_offset}, with the optical relaxation proxies tracing the centroid shift analagously to photon asymmetry.  The $w - A^2$ relationship has a best-fit positive slope at $2.8\sigma$  and the Spearman test gives a positive correlation at the 99.9  per cent level.  The $w - M_2/M_1$ again shows a significant correlation as well with a positive slope at $3.6\sigma$  and a positive Spearman correlation at the 97  per cent level.  Finally, we find no evidence for a correlation between the centroid shift and the MMG offset, with a power-law slope consistent with zero and a Spearman p-value of 0.62 .
\par
Based on the results from this section we conclude that Anderson-Darling statistic provides the best correlation with X-ray asymmetry among the three optical relaxation measures, as it shows the strongest Spearman correlations with the X-ray relaxation proxies, and the $A_\mathrm{phot} - A^2$ and $w - A^2$ relationships have positive power-law slopes at $\gtrsim 3\sigma$.  Modulo scatter, this correlation lends credence to the use of the Anderson-Darling test to quantify cluster relaxation for a large sample, as the shape of the diffuse X-ray profile is an independent (and arguably more direct) probe of the degree to which groups are unrelaxed/disturbed.  Therefore, for the remainder of this paper we will focus on the $A_\mathrm{phot} - A^2$ and $w - A^2$ relationships.  In the next section, we extend this analysis by investgating the halo mass dependence of the $A_\mathrm{phot} - A^2$ and $w - A^2$ relations.

\subsection{Halo mass dependence of X-ray-optical relations}
\label{sec:halo_mass}

In Section~\ref{sec:xray_opt_corr} we presented a significant correlation between the AD statistic for a given cluster and X-ray relaxation parameters (the photon asymmetry and the centroid shift).  In this section we further divide the sample into systems with small halo masses and those with large halo masses to investigate if the $A_\mathrm{phot} - A^2$ and $w - A^2$ correlations vary with cluster halo mass.  We choose the median halo mass of our cluster sample, $M_{\mathrm{halo},\,\mathrm{med}} = 10^{14.5}\Msun$, to make this division.

\subsubsection{Photon Asymmetry}
\label{sec:mh_aphot}

\begin{figure}
	\centering
	\includegraphics[width=0.9\columnwidth]{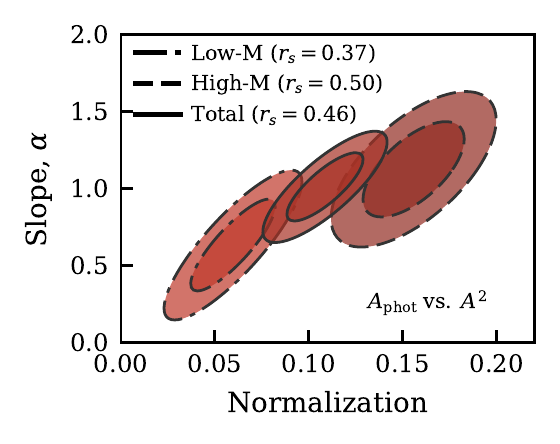}
	\caption{68 and 95 per cent confidence ellipses for the photon asymmetry vs. AD statistic best-fit power law parameters for low-mass halos ($M_\mathrm{halo} < 10^{14.5}\Msun$, dot-dashed), high-mass halos ($M_\mathrm{halo} > 10^{14.5}\Msun$, dashed), and the total sample (solid).  Spearman correlation coefficients are denoted for each sample.}
	\label{fig:asym_mh}
\end{figure}

In Fig.~\ref{fig:asym_mh} we show the confidence ellipses (68 and 95 per cent levels) corresponding to the power-law fit results to the $A_\mathrm{phot} - A^2$ relationship for high-mass and low-mass halos (as well as the total sample).  The separation between the high- and low-mass ellipses suggests that high- and low-mass clusters follow somewhat different scaling relations between $A_\mathrm{phot}$ and $A^2$.  For high-mass halos a significant correlation is still seen, with a best-fit slope of $1.15_{-0.39}^{+0.42}$  and a Spearman correlation significant at the 99.4  per cent level.  For low-mass halos the correlation is somewhat weaker with a best-fit power law slope of $0.61_{-0.33}^{+0.44}$  and a Spearman correlation significant at the 94  per cent level.  The best-fit slopes for the low- and high-mass halos are equal within uncertainties, however the normalization is larger for high-mass halos at the $> 2\sigma$ level.  Additionally, we find that high-mass halos have a slightly larger median $A_\mathrm{phot}$ ($0.08 \pm 0.03$ ) than low-mass halos ($0.04 \pm 0.01$ ).

\subsubsection{Centroid shift}
\label{sec:mh_cs}

\begin{figure}
	\centering
	\includegraphics[width=0.9\columnwidth]{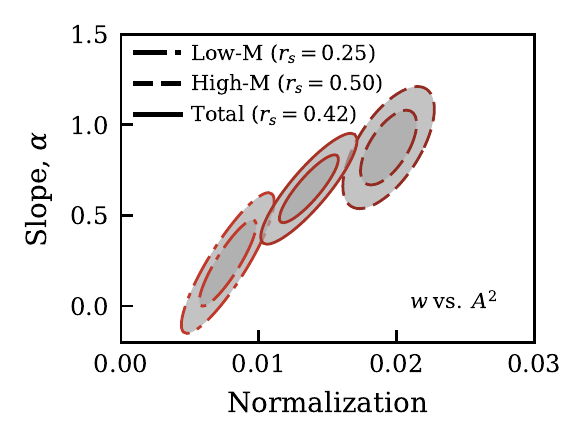}
	\caption{Same as Fig.~\ref{fig:asym_mh} however for the centroid shift instead of photon asymmetry.}
	\label{fig:cs_mh}
\end{figure}

In Fig.~\ref{fig:cs_mh} we now show the power-law fit results to the $w - A^2$ relationship for the two halo mass subsamples.  We find qualitatively similar results when considering centroid shift instead of photon asymmetry, with the high-mass halos displaying a clear correlation (slope: $0.86_{-0.24}^{+0.27}$ , Spearman p-value: 0.005 ) whereas the correlation for low-mass halos is marginal and not statistically significant (slope: $0.24_{-0.30}^{+0.35}$, Spearman p-value: 0.19 ).  We also find that the median centroid shift is larger for high-mass halos ($0.015 \pm 0.002$ ) than low-mass halos ($0.007 \pm 0.001$).

\subsection{The discrete case: X-ray asymmetry of Gaussian and non-Gaussian groups}
\label{sec:G_NG}

Thus far we have treated the shape of the velocity distribution in a continuous fashion with the AD statistic, though it is commonplace in the literature to define a dichotomy between ``Gaussian'' and ``non-Gaussian'' clusters \citep{hou2009, ribeiro2013a, roberts2017, decarvalho2017}.  We use the AD test and choose a critical p-value of 0.10 to define G and NG groups -- where G groups have $p_{AD} \ge 0.10$ and NG groups have $p_{AD} < 0.10$ (though our results are not sensitive to the precise p-value chosen over a reasonable range).
\par
To quantify the relationship between photon asymmetry and whether a cluster is classified as G or NG we employ the method of logistic regression \citep[e.g.][]{cox1958}.  Logistic regression is a classification tool used to estimate the probability of a binary response as a function of one (or many) independent variables, which may be numeric or categorical.  For this application, a galaxy cluster is classified as either G or NG (the boolean, dependent variable) and we are interested in the probability of a galaxy cluster being NG as a function of photon asymmetry or centroid shift (the numeric, independent variable).  The estimated probability is then

\begin{equation} \label{eq:log_curve}
  \hat{p} = \frac{e^{\beta_1 x + \beta_0}}{1 + e^{\beta_1 x + \beta_0}},
\end{equation}
\noindent
where $\beta_0$ and $\beta_1$ are parameters of the fit, and for this work we have $\hat{p} = \hat{p}(\mathrm{NG})$ and $x = \log A_\mathrm{phot}\;\mathrm{or}\; \log w$.

\subsubsection{Photon asymmetry}
\label{sec:G_NG_aphot}

\begin{figure}
	\centering
	\includegraphics[width=0.9\columnwidth]{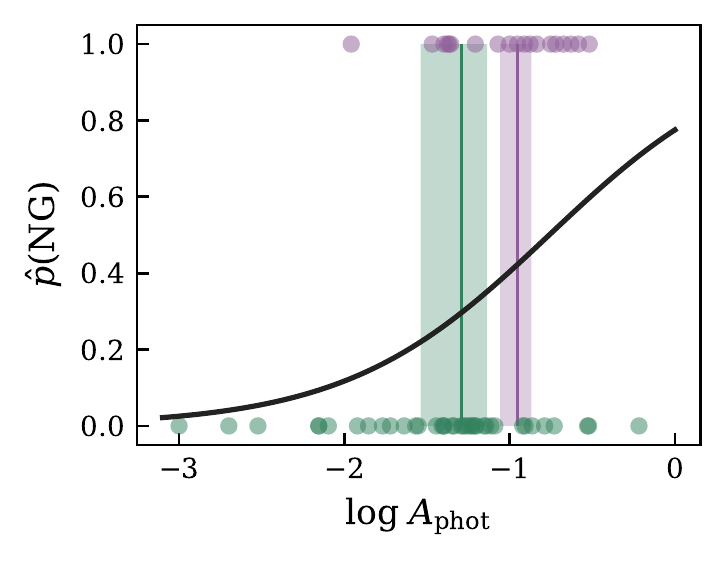}
	\caption{Estimated probability of a cluster being non-Gaussian as a function of photon asymmetry.  Photon asymmetry data points are shown for Gaussian (0, green) and non-Gaussian (1, purple) clusters and the black line shows the best-fit logistic curve.  Shaded vertical lines show the median asymmetry and $1\sigma$ standard error for Gaussian and non-Gaussian groups.}
	\label{fig:log_reg}
\end{figure}

In Fig.~\ref{fig:log_reg} we show the photon asymmetry for G (0, green) and NG (1, purple) clusters along with the best-fitting logistic curve (black line, equation~\ref{eq:log_curve}) describing the probability of being classified as NG as a function of $A_\mathrm{phot}$.  It is clear from Fig.~\ref{fig:log_reg} that the probability of a cluster being NG increases with photon asymmetry, we obtain a best-fit coefficient of $\beta_1 = 2.1 \pm 0.8$ indicating a significant correlation at $2.6\sigma$.  According to our logistic model, the $A_\mathrm{phot}$ value where the probability of being a NG cluster reaches 50 per cent is $A_\mathrm{phot} = 0.14$ and the asymmetry where the probability reaches 75 per cent is $A_\mathrm{phot} = 0.46$.  Additionally in Fig.~\ref{fig:log_reg} we show the median photon asymmetry and the $1\sigma$ standard error for G and NG clusters, NG clusters have a larger median asymmetry of $0.13 \pm 0.03$ compared to $0.05 \pm 0.02$ for G clusters.

\subsubsection{Centroid shift}
\label{sec:G_NG_cs}

\begin{figure}
	\centering
	\includegraphics[width=0.9\columnwidth]{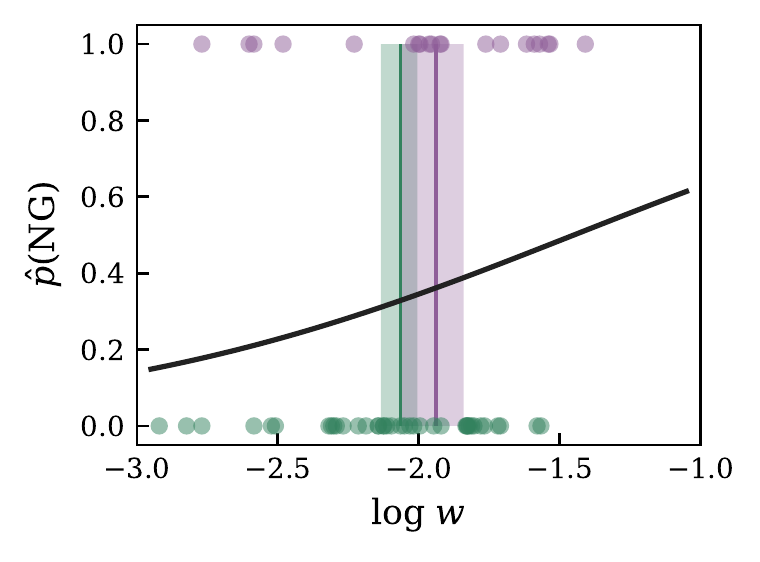}
	\caption{Same as Fig.~\ref{fig:log_reg} however for the centroid shift instead of photon asymmetry.}
	\label{fig:log_reg_w}
\end{figure}

In Fig.~\ref{fig:log_reg_w} we show an analagous logistic regression to Fig.~\ref{fig:log_reg}, with the centroid shift as the numeric variable.  Examining Fig.~\ref{fig:log_reg_w} it is clear that the distinction between G and NG clusters is not as strong as it was with photon asymmetry.  From the fit we obtain a best-fit coefficient of $\beta_1 = 1.2 \pm 0.8$, indicating only a marginal correlation at $1.5\sigma$.  From this fit the value of $w$ where $\hat{p}(\mathrm{NG})$ reaches 50 per cent is $w = 0.04$.  We also show the median value for $w$ for G (green) and NG (purple) clusters, and find that while the median centroid shift is slightly larger for NG clusters, this difference is not significant (G: $0.009 \pm 0.001$, NG: $0.012 \pm 0.003$).

\section{Discussion}
\label{sec:discussion}

\subsection{The Anderson-Darling test as a relaxation proxy}

The primary result from this paper is the strong correlation detected between X-ray relaxation measures and both the AD statistic for a given cluster (Fig.~\ref{fig:asym_ad_m12_offset}a and Fig.~\ref{fig:centshift}a), as well as the probability of a system being NG when considering the dichotomy of G and NG clusters (at least for $A_\mathrm{phot}$, Fig.~\ref{fig:log_reg}).  We argue that this is an important confirmation of the usefulness of the AD test to quantitatively identify unrelaxed/disturbed systems.  This, however, is only true in the statistical sense as there is still significant scatter around the $A_\mathrm{phot} - A^2$ and $w - A^2$ relations.  The AD test may or may not accurately classify the dynamical state of an individual cluster, but applied to a large statistical sample it is a useful tool to identify systems which are on average relaxed or unrelaxed.  It is also worth considering whether the group finder preferentially selects G or NG clusters.  The \citeauthor{yang2005} group finder constructs clusters assuming that the galaxy phase-space distribution follows a spherical NFW profile, which could bias the group finder in favour of G clusters (ie.\ assuming a spherical, symmetric distribution).  The analysis presented here does not account for any such bias, but since the clusters are all selected with the same algorithm the correlations found are robust for this sample.
\par
The AD test has become a relatively common tool used to identify unrelaxed systems from large redshift surveys \citep[e.g.][]{hou2009, ribeiro2010, martinez2012, hou2013, ribeiro2013a, roberts2017}, though its efficacy has only been tested in detail using Monte Carlo simulations sampling from idealized parent distributions (both Gaussian and non-Gaussian, \citealt{hou2009, ribeiro2013a}).  These tests have provided useful insight into the strengths and limitations of the AD test, but it is also important to test this technique in a more physical setting.  The comparision to diffuse X-ray morphology in this work provides one such test in an astronomical context.  In an upcoming paper we perform a detailed analysis on the AD test applied to groups and clusters in large, cosmological, N-body simulations.  This will allow us to explore things such as the false-positive rate, as well as potential differences in satellite time-since-infall or halo age for G and NG systems in a cosmological context.
\par
The results of Section~\ref{sec:G_NG} can also be used to constrain the dividing line between relaxed and unrelaxed clusters.  Based on the logistic regression model, the value of $A_\mathrm{phot}$ above which the probability of being a NG cluster exceeds 50 per cent is $A_\mathrm{phot} = 0.14$ and the value above which the probability exceeds 75 per cent is $A_\mathrm{phot} = 0.46$.  Correspondingly, the median $A_\mathrm{phot}$ for NG clusters of $0.13 \pm 0.03$, suggesting that $A_\mathrm{phot} \gtrsim 0.10 - 0.50$ may be a useful dividing line between relaxed and unrelaxed clusters, depending on the desired level of purity.  In \citet{mcdonald2017} a threshold of $A_\mathrm{phot} < 0.10$ is chosen to identify relaxed clusters, while this threshold was chosen arbitrarily we've shown here that this is a reasonable choice based on cluster velocity distribution measurements.  The threshold used to identify unrelaxed clusters in \citet{mcdonald2017} is $A_\mathrm{phot} > 0.50$, which also shows excellent agreement with the dividing lines that we derive from velocity measurements.  The choice of $A_\mathrm{phot} > 0.50$ is motivated by simulations of cluster major mergers from \citet{nurgaliev2017} who suggest that $A_\mathrm{phot} \gtrsim 0.2 - 0.6$ is a useful threshold to identify disturbed clusters, again corresponding very closely to the range we determine in this work.  This shows that using the AD test (in this case with a p-value of 0.10) to identify relaxed and unrelaxed clusters corresponds very closely to previous results using X-ray techniques.
\par
When using the centroid shift, $w$, instead of photon asymmetry, the logistic regression model does not seperate G and NG clusters as distinctly, however we can still use the model to constrain a dividing line.  In particular, the regression model suggests that the probability of being a NG cluster reaches 50 per cent at $w = 0.040$.  This is larger by a factor of a few than the boundary between regular and disturbed objects in previous X-ray analyses, which ranges between $w \simeq 0.01$ and $w \simeq 0.02$ \citep{ohara2006, cassano2010, weissmann2013}.  Given that the logistic regression only detects a marginal correlation between $\hat{p}(\mathrm{NG})$ and $w$, the dividing line that we derive here is likely not be well constrained.
\par
We can also contrast the two different X-ray relaxation proxies by highlighting any differences in the photon asymmetry and centroid shift relationships with the AD statistic.  When considering the continuous case in Section~\ref{sec:xray_opt_corr} we see very similar behaviour in the $A_\mathrm{phot} - A^2$ and $w - A^2$ relationships, consistent with the fact that photon asymmetry and centroid shift have been shown to correlate strongly \citep{nurgaliev2013}.  However the discrete case in Section~\ref{sec:G_NG} shows that G and NG clusters are more clearly segregated in terms of photon asymmetry than centroid shift, perhaps suggesting that photon asymmetry is a slightly stronger identifier of dynamically unrelaxed clusters, though a larger sample is required to robustly determine this.
\par
Finally, in Section~\ref{sec:halo_mass} we explored the halo mass dependence of the $A_\mathrm{phot} - A^2$ and $w - A^2$ relationships by separating the sample into subsamples of low-mass ($M_\mathrm{halo} < 10^{14.5}\Msun$) and high-mass clusters ($M_\mathrm{halo} \ge 10^{14.5}\Msun$).  We find small differences between the low-mass and high-mass relations, namely, both the $A_\mathrm{phot} - A^2$ and $w - A^2$ relationships for high-mass clusters have larger normalizations, whereas the slopes are consistent between the high- and low-mass samples.  In addition, the median values for $A_\mathrm{phot}$ and $w$ are slightly larger for high-mass clusters compared to low-mass clusters.  \citet{nurgaliev2013} show that the $A_\mathrm{phot}$ and $w$ statistics are robust against varying numbers of X-ray counts above $\sim 2000$ counts ($A_\mathrm{phot}$ is robust even below 2000 counts).  All of the systems in this work have $N_\mathrm{counts} > 2000$, therefore it is unlikely that these differences in asymmetry and centroid shift are being driven by the relatively high-count observations of massive systems.  This result hints that low-mass and high-mass clusters may follow slightly different scaling relations when it comes to $A_\mathrm{phot}$ or $w$ versus $A^2$, though a larger sample is necessary to build up the statistics required to conclude this with high confidence.  In principle, this difference could be explained through simple hierarchical growth where low-mass halos are on average more virialized than higher-mass clusters at the present day.  High-mass clusters will be more recently formed through mergers and accretion which can in turn increase $A_\mathrm{phot}$ and $w$ \citep[e.g.][]{cassano2010, nurgaliev2017}.  From the optical perspective, we also find that high-mass clusters have velocity distributions which are less Gaussian than low-mass systems, in agreement with previous studies \citep{roberts2017, decarvalho2017}.  Although it is important to note that it is easier to statistically identify departures from normality for high-mass systems with many members.

\subsection{Interpreting MMG-based relaxation parameters}
\label{sec:mmg_discussion}

The second optical relaxation proxy that shows a significant correlation with X-ray relaxation proxies is the stellar mass gap between the two most-massive cluster galaxies (see Section~\ref{sec:Aphot_corr} and Section~\ref{sec:w_corr}).  The correlations between X-ray relaxation proxies and $M_2 / M_1$ are found to be weaker than for $A^2$ (especially as measured by the Spearman correlation), potentially suggesting that $M_2 / M_1$ is a poorer (though still useful) tracer of diffuse X-ray morphology.  This may be expected given that satellite galaxies (i.e.\ the velocity distribution) and the diffuse hot gas profile should both trace the larger-scale cluster potential relatively directly, whereas central galaxy growth is governed more by dynamical interactions and gas accretion at the cluster centre.  It is also possible that the $A_\mathrm{phot} - M_2 / M_1$ and $w - M_2 / M_1$ trends are being affected by selection effects related to the difficulty indentifying the true MMG (and second most massive galaxy) in these clusters.  A particular concern regarding the SDSS is the impact of fibre collisions in the dense inner regions of clusters, as it has been estimated that up to 30 per cent of clusters may be missing a spectra for the true BCG \citep{vonderlinden2007}.  In an attempt to mitigate the effect of fibre collisions we use the systems from sample III in the Yang group catalogue which attaches redshifts to galaxies that lack spectra due to fibre collisions by assigning these galaxies the redshift of the galaxy it ``collided'' with.  While this procedure allows the group finder to include galaxies which otherwise would be missed due to fibre collisions, the trade-off is uncertainty regarding whether the added galaxies are true group members.  $\sim 60$ per cent of fibre collision galaxies have redshifts within $500\,\mathrm{km}\,\mathrm{s^{-1}}$ of the estimated value \citep{zehavi2002}, though this still leaves a significant number of fibre collision galaxies which may have true redshifts that differ substantially from the assigned value.  To ensure that our results are not being affected by the inclusion of these fibre collision galaxies we re-test the $A_\mathrm{phot} - M_2 / M_1$, $w - M_2 / M_1$ and $A_\mathrm{phot} - \delta R_{MMG}$, $w - \delta R_{MMG}$ relationships for correlations, now removing any systems where the MMG (and in the case of $M_2 / M_1$, the second-most-massive galaxy as well) is a fibre collision galaxy.  26 per cent of the clusters in the sample have an MMG which is a fibre collision galaxy and 39 per cent of the sample have either the MMG or the second-most-massive galaxy as a fibre collision galaxy.  Re-testing these relationships for correlations leaves the Spearman correlation coefficient virtually unchanged from Sections~\ref{sec:Aphot_corr} and \ref{sec:w_corr}, suggesting that fibre collision galaxies are not biasing the results.
\par
In Section~\ref{sec:Aphot_corr} we find no evidence for a correlation between $A_\mathrm{phot}$ or $w$ and $\delta R_{MMG}$, which suggests that the MMG offset is not a reliable tracer of cluster relaxation.  One caveat which is important to consider is the assumptions made to justify the use of $M_2 / M_1$ and $\delta R_{MMG}$ as relaxation proxies, in particular that for relaxed systems the MMG (or brightest galaxy) resides at rest at the centre of the dark matter potential -- the so-called central galaxy paradigm (CGP).  For example, if the MMG is instead a satellite galaxy then the use of $M_2 / M_1$ as a relaxation measure may not be valid as it is predicated on the MMG being the central and growing through accretion and mergers at the centre of the potential well.  Similarly, if the MMG is a satellite then its offset from the luminosity-weighted centre would not be expected to trace cluster relaxation.  Many recent studies have called into question the ubiquity of the CGP by highlighting the fact that a substantial fraction of brightest cluster galaxies (BCGs) are significantly offset from the cluster centroid, both in terms of projected distance and velocity \citep{vandenbosch2005, coziol2009, skibba2011, sehgal2013, lauer2014, hoshino2015}.  In particular, \citet{skibba2011} find that the fraction of halos where the brightest galaxy is in fact a satellite ($f_\mathrm{BNC}$) ranges from $\sim 25$ per cent in low-mass halos ($10^{12}\,h^{-1} \le M \lesssim 2 \times 10^{13}\,h^{-1}\Msun$) to $\sim 40$ per cent in high-mass halos ($M \gtrsim 5 \times 10^{13}\,h^{-1}\Msun$).  Furthermore, \citet{hoshino2015} find $f_\mathrm{BNC} \sim 20 - 30$ per cent for galaxies in redMaPPer clusters, and in terms of velocity \citet{coziol2009} show that the median peculiar velocity for BCGs in a sample of Abell clusters is $\sim$ one third of the cluster velocity dispersion.  It is plausible that systems where the CGP is not valid are diluting stronger trends between $A_\mathrm{phot}$ or $w$ and $M_2/M_1$, or perhaps masking trends between $A_\mathrm{phot}$ or $w$ and $\delta R_{MMG}$.  Unfortunately, identifying systems where the CGP is violated is difficult on a case-by-case basis, limited by observing in projection, and is generally done in the statistical sense for large samples (ie.\ thousands) of groups and clusters \citep[e.g.][]{vandenbosch2005, skibba2011}.  Therefore we continue to argue that the AD test (or some other measure of the velocity distribution shape, see e.g. \citealt{ribeiro2013a}) is a better optical relaxation proxy as it is not complicated by CGP assumptions.
\section{Summary}
\label{sec:summary}
In this paper we present a comparison between diffuse X-ray morphology and cluster relaxation proxies based on optical measures.  Using the \citet{yang2007} SDSS group catalogue we match optically identified clusters with $N \ge 10$ members to X-ray observations from both the \textit{Chandra} and \textit{XMM-Newton} X-ray observatories.  With a sample of 58 X-ray matched clusters we compare X-ray asymmetry and centroid shift to three different optical relaxation probes: the Anderson-Darling statistic, the stellar mass gap, and the MMG offset.  The main conclusions of this work are as follows:

\begin{enumerate}
\item We detect a significant positive correlation between X-ray relaxation proxies (photon asymmetry, centroid shift) and Anderson-Darling statistic at $\sim 3-4\sigma$ as measured by both a power-law fit and by the Spearman correlation test, and a weaker correlation ($\sim 2-3 \sigma$) between X-ray relaxation proxies and stellar mass gap (between two most-massive cluster galaxies).

\item We do not detect a significant correlation between X-ray asymmetry or centroid shift and the MMG offset.

\item We find that the $A_\mathrm{phot} - A^2$ and $w - A^2$ relationships vary somewhat for low-mass ($M_\mathrm{halo} < 10^{14.5}\Msun$) and high-mass ($M_\mathrm{halo} \ge 10^{14.5}\Msun$) clusters.  Specifically, high-mass clusters have a best-fit relationship with a larger normalization, and the median asymmetry and centroid shift is larger in high-mass systems.  However, a definitive measure of the halo mass dependence awaits a larger sample.

\item When considering a dichotomy between Gaussian ($p_{AD} \ge 0.10$) and non-Gaussian ($p_{AD} < 0.10$) clusters we find that the probability of being a non-Gaussian system (as measured by a logistic regression) correlates clearly with X-ray asymmetry.  Additionally, the median asymmetry of non-Gaussian clusters is larger than that of Gaussian clusters.  When using the centroid shift as the X-ray relaxation proxy the correlation is marginal.
\end{enumerate}
\noindent
Though the scatter in the above mentioned relations limit the reliability of this approach on a case-by-case basis, these results confirm the effectiveness of the shape of the projected velocity distribution as a proxy for cluster relaxation, when applied to a large sample.


\section*{Acknowledgments}
\label{sec:acknowledgments}
We thank the Natural Sciences and Engineering Research Council of Canada for funding.  This work made use of many open-source software packages, such as: \textsc{AstroPy} \citep{astropy2013}, \textsc{Matplotlib} \citep{hunter2007}, \textsc{NumPy} \citep{vanderwalt2011}, \textsc{Pandas} \citep{mckinney2010}, \textsc{Photutils} \citep{bradley2016}, \textsc{SciPy} \citep{jones2001}, \textsc{Statsmodels} \citep{seabold2010}, and \textsc{Topcat} \citep{taylor2005}.
\par
Funding for the Sloan Digital Sky Survey IV has been provided by the Alfred P. Sloan Foundation, the U.S. Department of Energy Office of Science, and the Participating Institutions. SDSS-IV acknowledges
support and resources from the Center for High-Performance Computing at
the University of Utah. The SDSS web site is www.sdss.org.
\par
SDSS-IV is managed by the Astrophysical Research Consortium for the 
Participating Institutions of the SDSS Collaboration including the 
Brazilian Participation Group, the Carnegie Institution for Science, 
Carnegie Mellon University, the Chilean Participation Group, the French Participation Group, Harvard-Smithsonian Center for Astrophysics, 
Instituto de Astrof\'isica de Canarias, The Johns Hopkins University, 
Kavli Institute for the Physics and Mathematics of the Universe (IPMU) / 
University of Tokyo, Lawrence Berkeley National Laboratory, 
Leibniz Institut f\"ur Astrophysik Potsdam (AIP),  
Max-Planck-Institut f\"ur Astronomie (MPIA Heidelberg), 
Max-Planck-Institut f\"ur Astrophysik (MPA Garching), 
Max-Planck-Institut f\"ur Extraterrestrische Physik (MPE), 
National Astronomical Observatories of China, New Mexico State University, 
New York University, University of Notre Dame, 
Observat\'ario Nacional / MCTI, The Ohio State University, 
Pennsylvania State University, Shanghai Astronomical Observatory, 
United Kingdom Participation Group,
Universidad Nacional Aut\'onoma de M\'exico, University of Arizona, 
University of Colorado Boulder, University of Oxford, University of Portsmouth, 
University of Utah, University of Virginia, University of Washington, University of Wisconsin, 
Vanderbilt University, and Yale University.




\bibliographystyle{mnras}
\bibliography{CompleteManuscriptFile_v3.bib} 




\appendix

\section{X-ray images}

\begin{figure*}
	\centering
	\includegraphics[width=\textwidth]{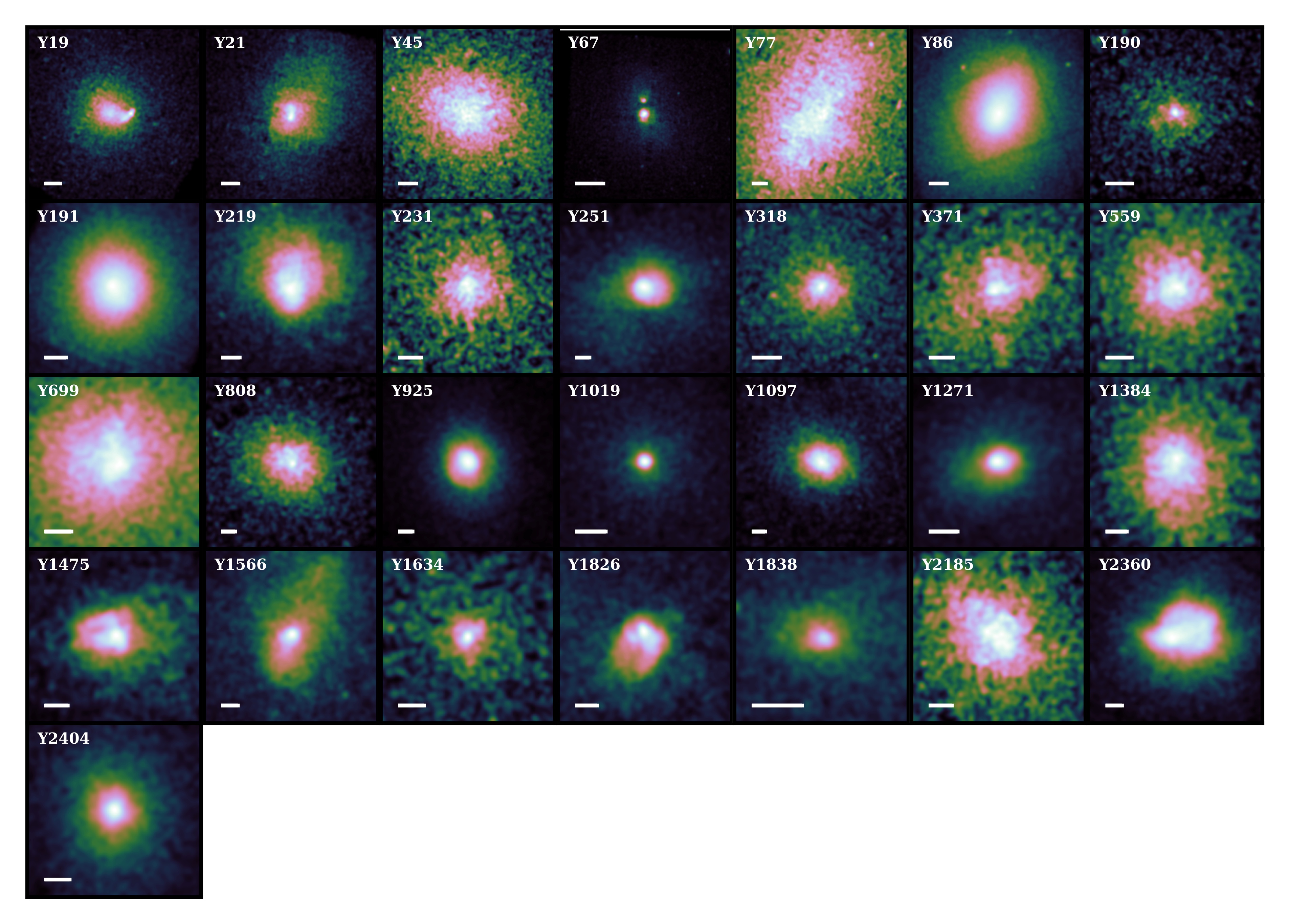}
	\caption{$0.5-5\,\mathrm{keV}$ images for the clusters in our sample which are observed by \textit{Chandra}.  Thumbnail have dimensions of $R_{500} \times R_{500}$ and are smoothed using a Gaussian filter with a bandwidth of $5\arcsec$.  The scalebar in each image corresponds to a physical size of $100\,\mathrm{kpc}$.}
	\label{fig:images_chandra}
\end{figure*}

\begin{figure*}
	\centering
	\includegraphics[width=\textwidth]{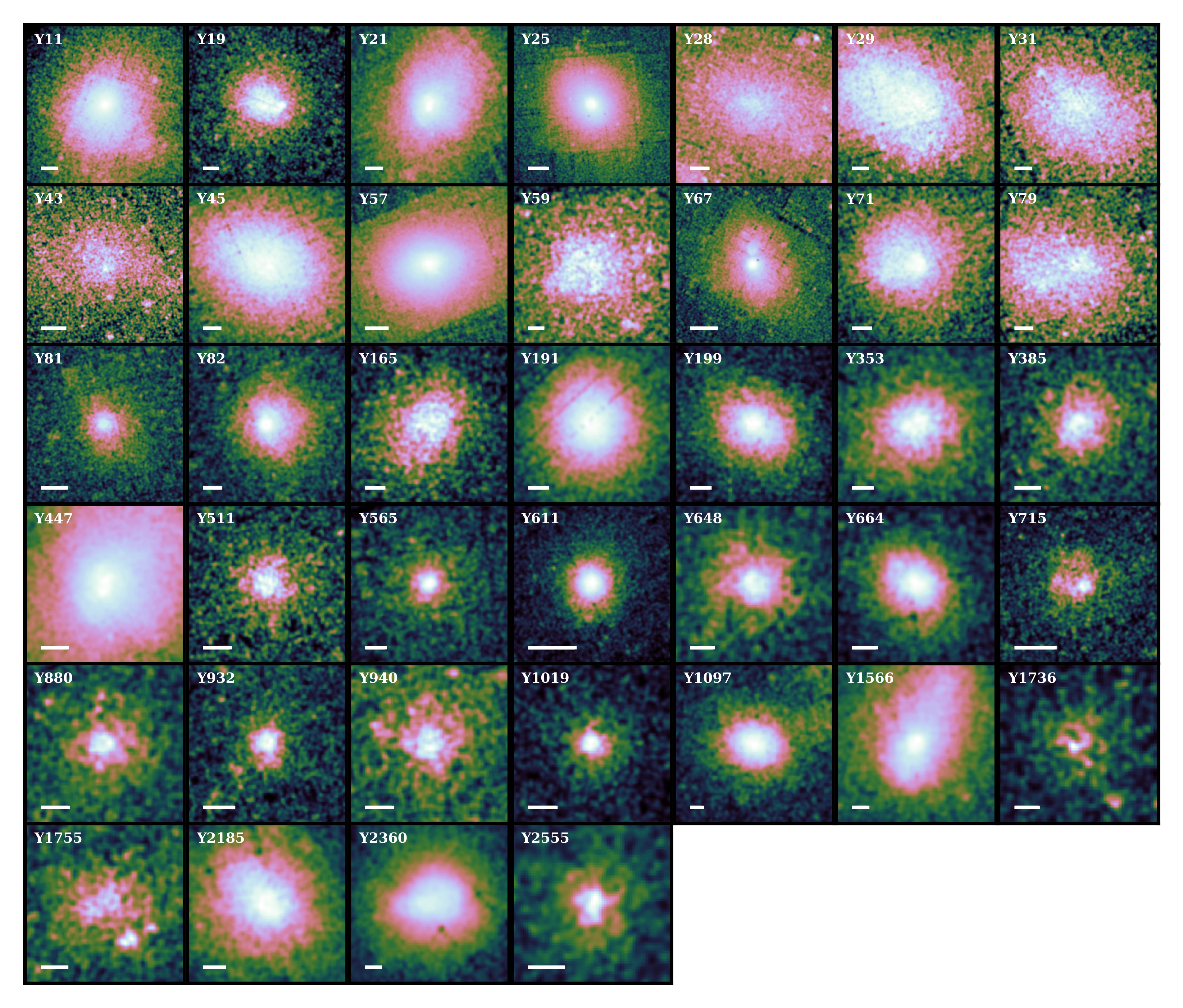}
	\caption{Same as Fig.~\ref{fig:images_chandra} but for \textit{XMM-Newton} observations.}
	\label{fig:images_xmm}
\end{figure*}

\bsp	
\label{lastpage}
\end{document}